\documentclass[%
 reprint,
 amsmath,amssymb,
 aps,
]{revtex4-2}

\usepackage{graphicx}
\usepackage{dcolumn}
\usepackage{bm}
\usepackage{mathrsfs}
\usepackage[english]{babel}
\usepackage{amsthm}
\usepackage{bbm} 
\usepackage{float}
\usepackage{graphics}
\usepackage{comment}
\usepackage[inline]{enumitem}
\usepackage{xcolor}
\usepackage{bbm}

\newcommand{\jt}[1]{\textcolor{blue}{\small \ [[{\bf JT:}\it #1]]\ }}

\usepackage[normalem]{ulem}

\newcommand{\E}{\mathbb E}

\usepackage{xcolor}

\usepackage{subcaption}
\usepackage[colorlinks]{hyperref}
\usepackage[nameinlink]{cleveref} 

\crefname{figure}{Fig.}{Figs.}
\crefname{equation}{Eq.}{Eqs.}

\begin{document}

\preprint{APS/123-QED}

\title{Complexity and dynamics of partially symmetric random neural networks}

\author{Nimrod Sherf$^{1}$}
\email{nsherf@uh.edu }
\author{Si Tang$^{2}$}
\email{sit218@lehigh.edu }
\author{Dylan Hafner$^{2}$}
\email{djh320@lehigh.edu}
\author{Jonathan D. Touboul$^{3}$}
\email{jtouboul@brandeis.edu}
\author{Xaq Pitkow$^{4,5,6,7,8}$}%
\email{xaq@cmu.edu}
\author{Kevin E. Bassler$^{1,9,10}$}%
\email{bassler@uh.edu}
\author{Krešimir Josi\'{c}$^{1,11,12}$}%
\email{kresimir.josic@gmail.com}
\affiliation{%
$^1$Department of Mathematics, University of Houston, Houston, TX,
$^2$Department of Mathematics, Lehigh University, Bethlehem, PA,
$^3$ Department of Mathematics and Volen National Center for Complex Systems, Brandeis University, Waltham, MA, 
$^4$Neuroscience Institute, Carnegie Mellon University, Pittsburgh, PA,
$^5$Department of Machine Learning in the School of Computer Science, Carnegie Mellon University, Pittsburgh, PA,
$^6$Department of Neuroscience, Baylor College of Medicine, Houston, TX,
$^7$Departments of Electrical and Computer Engineering, and Computer Science, Rice University, Houston, TX,
$^8$NSF AI Institute for Artificial and Natural Intelligence,
$^9$Department of Physics, University of Houston, Houston, TX,
$^{10}$Texas Center for Superconductivity, University of Houston, Houston, TX,
$^{11}$Department of Biology and Biochemistry, University of Houston, Houston, TX,
$^{12}$NSF-Simons National Institute for Theory and Mathematics in Biology, Chicago, IL. \\
}%
\date{\today}

\begin{abstract}

Neural circuits exhibit structured connectivity, including an overrepresentation of reciprocal connections between neuron pairs. Despite important advances, a full understanding of how such partial symmetry in connectivity shapes neural dynamics remains elusive. Here we ask how correlations between reciprocal connections in a random, recurrent neural network affect phase-space complexity, defined as the exponential proliferation rate (with network size) of the number of fixed points that accompanies the transition to chaotic dynamics. We find a striking pattern: partial anti-symmetry strongly amplifies complexity, while partial symmetry suppresses it. These opposing trends closely track changes in other measures of dynamical behavior, such as dimensionality, Lyapunov exponents, and transient path length, supporting the view that fixed-point structure is a key determinant of network dynamics. 
Thus, positive reciprocal correlations favor low-dimensional, slowly varying activity, whereas negative correlations promote high-dimensional, rapidly fluctuating chaotic activity. These results yield testable predictions about the link between connection reciprocity, neural dynamics and function.

\end{abstract}

\maketitle


Synaptic connectivity is not random. Mounting experimental evidence shows that the structure of biological neural networks is complex and characterized by correlations between pairs, triplets, and even higher multiples of connectivity weights~\cite{white1986structure,milo2002network,sporns2004motifs,song2005highly,bullmore2009complex,perin2011synaptic,ko2011functional,levy2012spatial,duclos2021brain,yang2023cyclic}.  
Theoretical work has established that these complex connectivity patterns shape neuronal dynamics and function \cite{wang2006heterogeneity,marti2018correlations,mastrogiuseppe2018linking,rao2019dynamics,berlemont2022glassy,deshpande2023third,sherf2025chaos}. 
In particular, reciprocal connections between neuron pairs are overrepresented across multiple brain regions \cite{song2005highly,perin2011synaptic,larimer2008nonrandom,markram1997physiology,wang2006heterogeneity}, and are thought to influence memory formation \cite{larimer2008nonrandom,zaitsev2013functional}, learning \cite{cervantes2017reciprocal}, and information processing \cite{ko2011functional,cossell2015functional}. Despite their importance for function, the impact of pairwise synaptic weight correlations on neural dynamics is not fully understood.  

The relation between  connectivity structure and neural dynamics is often studied using simplified Recurrent Neural Network (RNN) models. This approach has provided  insights into the different dynamical regimes that characterize neural activity~\citep{sompolinsky1988chaos,mastrogiuseppe2018linking,landau2018coherent,van1996chaos}, and has played a role in the development of neuro-inspired models of machine learning~\cite{mante2013context,driscoll2024flexible}. Large, strongly coupled RNNs with unstructured, random connectivity  exhibit high-dimensional, chaotic dynamics akin to the activity observed in neural networks \textit{in vivo}~\citep{sompolinsky1988chaos}. 
Moreover, fluctuations in chaotic neural activity slow as the probability  of reciprocal connections increases \cite{marti2018correlations}.

In such networks, random field theory has provided an analytically tractable approach to quantifying the topological complexity of phase space defined as the rate of growth of equilibria with system size~\cite{wainrib2013topological}.  This approach  has been used to show that in strongly coupled RNNs with independent and identically distributed connections, the number of unstable fixed points increases exponentially with the number of neurons. This explosion in the number of fixed points accompanies the emergence of high-dimensional, chaotic activity \cite{sompolinsky1988chaos,clark2023dimension}, and is related to other aspects of neuronal activity \cite{stubenrauch2025fixed}. 

Here, we ask how topological complexity is affected by the presence of correlations in reciprocal connectivity weights, and how the resultant changes in complexity impact the dynamics of neural networks. We analytically compute the complexity of networks with partially   symmetric weights, and find a striking impact of correlations in connectivity strengths: negative correlations accelerate the explosion in the number of fixed points, while positive correlations suppress it. We show numerically that these trends are consistent with changes in the largest Lyapunov exponent, dimensionality, and transient path length, supporting the view that the scaling in fixed-point number is closely linked to other aspects of network dynamics.   

We consider a classical $N$-neuron recurrent rate network~\cite{amari1972characteristics,sompolinsky1988chaos,mastrogiuseppe2018linking} with the membrane potential of neuron $i\in\{1,\dots, N\}$, denoted $x_i(t)$, evolving according to the differential equation 
\begin{align}
\label{eq:nn}
\dot x_i = -x_i + \sum_{j=1}^N w_{ij}S(x_j).
\end{align}
Here $\mathbf W = (w_{ij})_{i,j=1}^N$ is an $N$-by-$N$ matrix of Gaussian synaptic weights satisfying
\begin{align*}
    \langle w_{ij}\rangle&=0,\quad 
    \langle w_{ij}w_{kl}\rangle=\frac{g^2 }{N}(\delta_{ik} \delta_{jl}+\tau \delta_{il} \delta_{jk}),
\end{align*} 
where $\delta_{ij}$ is the Kronecker delta ($\delta_{ij}=1$ if $i=j$ and $0$ otherwise).
The gain parameter, $g,$ determines the coupling strength, the correlation coefficient $\tau \in [-1,1]$ 
controls the symmetry of the connectivity matrix $\mathbf W$, and $\langle \cdots \rangle$ denotes an ensemble average.
The $1/N$ scaling ensures that inputs to each neuron remain $O(1)$ in the large-$N$ limit. The function $S$ is often chosen to be an odd, sigmoid function with maximal slope at the origin ($0\le S'(x)\le S'(0)=1$), representing a synaptic nonlinearity. For concreteness, we use $S(x)=\tanh(x)$ in our numerical simulations.

A classical result of random matrix theory states that, in the limit as $N\to\infty$, the eigenvalue distribution of $\mathbf W$ follows the elliptical law~\cite{girko1986elliptic,sommers1988spectrum,kuczala2016eigenvalue}, and approaches the uniform distribution on the ellipse 
\begin{align}
\label{eqn:ellipse}
    \mathcal E_{ab}:=\{u+iv\in \mathbb C:  u^2/a^2 + v^2/b^2\leq 1\}
\end{align}
with semi-axes $a=g(1+\tau)$ and $b=g(1-\tau)$.
The well-known \emph{circular law}  holds at $\tau=0$, where the limiting spectral distribution is  uniform  on the disk of radius $g$~\cite{girko1985circular}, while \emph{Wigner's semi-circle law}~\cite{Wigner1958} holds at $\tau = 1$ when $\mathbf W$ is symmetric and the spectral distribution collapses to the real line segment $[-2g, 2g]$.

Our  goal is to characterize the growth rate  in the number of fixed points, $A_N(g, \tau),$ of the neural network defined by \cref{eq:nn} as $N\to \infty$. Note that $A_N(g, \tau)$ is a random variable that depends on the realization of the entries in $\mathbf W$. Each fixed point is a vector, $\mathbf x^\ast = (x_1^\ast, \ldots, x_N^\ast),$ satisfying the system of  equations
\begin{align}
\label{eq:eq}
-x^\ast_i +\sum_{j=1}^{N} w_{ij} S(x^\ast_j)=0, \ \ \text{for }i = 1,2,\dots, N.
\end{align}
This system is generally difficult to solve both analytically and numerically. 

The \emph{topological complexity} is defined as the rate of exponential growth of the expected number of fixed points relative to the network size, that is,
$$
c(g,\tau) := \lim_{N \rightarrow \infty} \frac1N \log \mathbb{E}[A_N (g, \tau)].
$$
It has been shown that when $\tau = 0$, \emph{i.e.}, when weights are independent, for large networks the system defined by Eq.~\eqref{eq:nn} undergoes a phase transition at $g = 1$: The system has a unique fixed point at the origin when $g<1$, while the number of fixed points grows  exponentially with system size when $g>1$~\cite{wainrib2013topological}. Specifically, if $g<1$, $\mathbb{E}[A_N (g, 0)]\sim 1$ and thus $c(g,0)=0$; and if $g>1$, $c(g, 0)\sim (g-1)^2$ as $g \to 1^+$.

In general,  the trivial fixed point at the origin for  Eq.~\eqref{eq:nn} becomes unstable when the rightmost eigenvalue of the matrix $\mathbf W$ has a real part greater than $1$. When the eigenvalue spectrum follows the elliptical law, this instability occurs when $g(1+\tau) = 1$, and we thus define the \emph{effective gain} as $g_{\text{eff}} \equiv g(1+\tau)$, so that $g_{\text{eff}}=1$ corresponds to the combination of parameters $(g, \tau)$ for which the origin almost surely loses stability when $N\to \infty$. We focus on the regime near $g_{\text{eff}}=1$, and refer to it as the \emph{onset} of instability.
When $g_{\text{eff}}<1$, we can show that the system is contracting as $N\to\infty$ and $\mathbf x^\ast = \mathbf 0$ is the unique solution to \cref{eq:eq}. Hence, 
$c(g_{\text{eff}},\tau) = 0$ for all $g_{\text{eff}}<1$ and all $\tau \in [-1,1]$. To prove this claim, we show in Appendix \ref{sec:spectrumWD} that the rightmost eigenvalue of $\mathbf W \cdot \Delta(S'(\mathbf x^\ast))$ is asymptotically no larger than $g_{\text{eff}}$, and use this bound in Appendix \ref{sec:uniq-fix}  to conclude that the fixed point at the origin is unique.

When $g_{\text{eff}}=1+\epsilon > 1$ for some small $\epsilon>0$,  following~\cite{wainrib2013topological}, we can show that for large $N$ all fixed points of \cref{eq:nn} are confined to the vicinity of the origin. To show this, we consider a perturbation of \cref{eq:nn} with $S_{\epsilon}$ replacing $S$. Choose $S_{\epsilon}$ as  a smooth modification of the function $S$ so that $0<S_\epsilon'(x) \le 1/g_{\text{eff}}$. This can be achieved by, for example, defining $S_{\epsilon}(x)=x/g_{\text{eff}}$ whenever $S'(x)>1/g_{\text{eff}}$. Since $\epsilon$ is small, this  perturbation is only needed in a ball $B_{r(\epsilon)}$ centered at the origin of a radius $r(\epsilon)$ with $r(\epsilon)\to 0$ as $\epsilon \to0$. The perturbed system has a unique fixed point at the origin, which can be shown using an argument similar to the one we used in the case $g_{\text{eff}}<1$. It follows that the original system given by \cref{eq:nn} can only have fixed points within the ball $B_{r(\epsilon)}$, as the dynamics of the perturbed and unperturbed systems outside $B_{r(\epsilon)}$ are identical. 

To count the expected number of fixed points inside the ball $B_{r(\epsilon)}$, we use the Kac-Rice formula \cite{kac1943average,wainrib2013topological,stubenrauch2025fixed} 
\begin{align}
    \notag
    \E [A_N(g, \tau)] = \int_{B_{r(\epsilon)}}\!\E\bigg [&\left | \text{det}\left (-\mathbf I + \mathbf W \cdot \Delta(S'(\mathbf x)) \right )\right| \cdot \\
    \label{eq:kacrice-1}
    &\delta_0 (- \mathbf{x} + \mathbf{W} \cdot S(\mathbf{x})) \bigg ] \ \mathrm{d} \mathbf{x}.
\end{align}
From the continuity of the determinant function it follows that as $\epsilon \to 0$, 
the integral in \cref{eq:kacrice-1} approaches
\begin{align}
\label{eqn:det-approx}
    \E [A_N(g, \tau)] &=\E [|\text{det}(-\mathbf I + \mathbf W)|]\big(1 + \eta_1(N, \epsilon)\big).
\end{align}
where $1/N\log \eta_1(N, \epsilon)\to 0$ as $N\to \infty$. 
The determinant $\E [|\text{det}(-\mathbf I + \mathbf W)|]$ can be approximated in the limit $N\to\infty$ as 
\begin{align*}
    &\frac{1}{N}\log\E [|\text{det}(-\mathbf I + \mathbf W)|]\approx \frac{1}{N}\E \!\!\sum_{\lambda \in \text{Sp}(\mathbf W)}^N\!\log |-1+\lambda|\\
    &\quad \qquad \to \int_\mathbb C \log|-1+z|\,  \rho_{g, \tau}(z)\mathrm{d}z\, \text{ as } N \to \infty,
\end{align*}
where $\rho_{g, \tau}$ is the density of the uniform distribution on the ellipse $\mathcal E_{ab}$, see \cref{eqn:ellipse}. Evaluating the integral (see details in Appendix \ref{sec:topo-complex}), we obtain that 
\begin{align*}
c(g, \tau) &=
\frac{1}{2g^2(1+\tau)}-\frac{1}{2}+\log g.
\end{align*}

At the onset as $g_{\text{eff}}\to 1^+$, the expansion  of the topological complexity to the second order in $(g_{\text{eff}}-1)$ is given by: 
\begin{align}  
\notag
c(g_\mathrm{eff}, \tau)\approx & 
-(g_\mathrm{eff}-1) \tau + (g_\mathrm{eff}-1)^2 \left (\frac{3}{2} \tau +1 \right) \\
\label{eq:TC}
&+\frac{\tau }{2}-\log \left(1+\tau\right).
\end{align}
This observation thus implies that the uncorrelated weight matrices ($\tau=0$) mark a transition between two vastly distinct regimes corresponding to positively and negatively correlated weights. For $\tau=0$, the complexity displays a second-order phase transition, with a  complexity increasing  quadratically with $g_\mathrm{eff}$, implying that the edge of chaos is of order $O(N^{-1/2})$~\cite{wainrib2013topological}. For $\tau \neq 0$ and  $g_{\mathrm{eff}}$ close to unity,  changes in $c(g_{\text{eff}},\tau)$ are instead dominated by the linear term $-\tau (g_{\mathrm{eff}}-1)$. This leads to fundamentally different regimes depending on the sign of $\tau$, that is, on whether reciprocal weights are positively or negatively correlated.

When $\tau<0$, the topological complexity, $c(g_{\mathrm{eff}}, \tau)$ is positive and increases linearly with $g_{\mathrm{eff}}$ near the onset. Thus, negative reciprocal weight correlations lead to a much faster exponential rate of  growth in the number of fixed points  compared to random networks with unstructured weight distributions, exhibiting a first-order phase transition. In particular, the edge of chaos is now of order $O(N^{-1})$, implying that bifurcations leading to the emergence of fixed points occur within a much smaller parameter interval.

\begin{figure}[ht]
    \centering
    \includegraphics[width=0.40\textwidth]{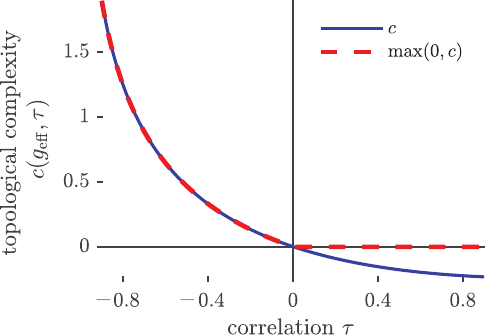}
    \caption{Topological complexity, $c,$ given in \cref{eq:TC} as a function of the symmetry parameter, $\tau,$ for $g_{\mathrm{eff}}=1.05$  (solid blue line). The dashed red line shows $\max(0,c)$, the contribution of $c$ to the scaling of the expected number of fixed points $\E[A_N]$ in the limit $N \rightarrow\infty$.}
     \label{fig:fig1}
\end{figure}

At onset, when $\tau \gtrsim 2(g_{\mathrm{eff}}-1)^2$, the right-hand side of \cref{eq:TC} is negative. Consequently, for $g_{\mathrm{eff}}$ sufficiently close to one and positively correlated reciprocal weights, the contribution of $c(g_{\mathrm{eff}},\tau)$ to the exponential proliferation of fixed points vanishes as $N\to\infty$ (see \cref{fig:fig1}). Thus, the complexity may not be fully captured by our exponential estimate and may instead be governed by the $\eta_1(N,\epsilon)$ term, which we do not estimate here. Henceforth, we refer to $c(g_{\mathrm{eff}},\tau)$ as the topological complexity, while emphasizing that for $\tau>0$ the true complexity of phase space is not described by \cref{eq:TC}.

Fixing $g_{\mathrm{eff}}$ and observing the effects of correlations at the onset shows  that topological complexity decreases monotonically as $\tau$ increases (see \cref{fig:fig1}). Furthermore, the limit of  completely anti-symmetric coupling,  $\tau\rightarrow-1,$ is singular, and the topological complexity diverges: the landscape shifts from exponentially many fixed points for $\tau>-1$ to a single fixed point at $\tau=-1$, where the spectrum of the connectivity matrix is purely imaginary.


\begin{figure}[ht]
    \centering
    \includegraphics[width=0.40\textwidth]{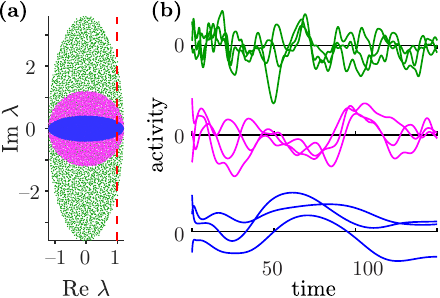}
    \caption{(a) The distributions of eigenvalues of the connectivity matrix $\mathbf{W}$ for $\tau=-0.5$ (green), $\tau=0$ (magenta), and $\tau=0.5$ (blue) with effective gain $g_{\text{eff}}=1.2$, and $N=4000$. The dashed red line corresponds to  $\mathrm{Re}\,\lambda=1$. (b)  Three representative solutions of the corresponding system given by \cref{eq:nn}. The colors of the trajectories match the colors of the connectivity matrices in panel (a).  \label{fig:fig2}} 
\end{figure}

To what extent do the dynamics of partially symmetric networks reflect the topological complexity? Recent work shows that increasing symmetry slows the characteristic timescale of chaotic fluctuations~\cite{marti2018correlations}. A heuristic explanation follows from observing that increasing $\tau$ for a fixed $g_{\mathrm{eff}}$ compresses the spectrum of $\mathbf W$  in the vertical direction. Thus, the unstable eigenvalues of the Jacobian matrix evaluated at the origin approach the real axis and have decreasing imaginary parts, consistent with slower fluctuations (see \cref{fig:fig2}). Moreover, at $\tau=1$ the dynamics are relaxational and admit an energy function \cite{crisanti2018path}, implying convergence to a stable fixed point at long times.
In contrast, as $\tau$ decreases, fluctuations become faster, consistent with an increase of the imaginary parts of the leading eigenvalues of the Jacobian evaluated at the origin  (see \cref{fig:fig2}). Moreover, as $\tau \to -1$, the eigenvalue distribution of $\mathbf{W}$ stretches along the imaginary axis with increasingly long tails. This follows because, at fixed $g_{\mathrm{eff}}$, decreasing $\tau$ entails rescaling the coupling $g$ as $(1+\tau)^{-1}$. The resulting increase in $g$  amplifies high-frequency modes that shape the dynamics.

RNNs with changing  reciprocal weight correlations thus provide us with a simple model whose topological complexity varies continuously with a single parameter. This allows us to  test whether  topological complexity is related to  other measures of the network's dynamics, as conjectured previously for $\tau=0$~\cite{wainrib2013topological} where it was shown that  complexity scales as the maximal Lyapunov exponent~\cite{wainrib2013topological,sompolinsky1988chaos,crisanti2018path,stubenrauch2025fixed}. For such networks the phase transition in the number of fixed points also coincides with  the emergence of chaotic dynamics~\cite{sompolinsky1988chaos}. Moreover, it was shown that fixed points and typical trajectories concentrate on distinct shells in phase space, where a “shell” refers to the set of states with approximately the same population activity norm. Close to onset, the characteristic shells containing the fixed points and the attractors lie near one another, but they separate as $g$ increases \cite{stubenrauch2025fixed}. We therefore expect the relationship between complexity and activity to be strongest near onset, and increasingly indirect farther from onset. 

We next ask whether a similar correspondence arises between topological complexity and other metrics characterizing the neuronal dynamics.
It has been suggested that the unstable manifolds of the fixed points guide the dynamics of the system, so their geometry constrains the effective degrees of freedom that trajectories can explore~\cite{cessac1994mean,wainrib2013topological,stubenrauch2025fixed,yang2025relationship}. We thus expect that weight-correlation-induced changes in complexity should be reflected in the maximal Lyapunov exponent, $\lambda_{\mathrm{LE}},$ and in the dimensionality of activity. Consistent with this expectation, for $\tau \gtrsim -0.7$ the Lyapunov exponent decreases as $\tau$ increases, in parallel with a decrease in complexity (\hyperlink{Fig3}{Fig.\ 3a}). However, for sufficiently strong anti-symmetry, $\tau \lesssim -0.7$, $\lambda_{\mathrm{LE}}$ drops sharply even as the complexity continues to rise. This mismatch indicates that (\textit{i}) complexity is not the sole determinant of the largest Lyapunov exponent, $\lambda_{\mathrm{LE}}$, and (\textit{ii}) the fact that both the topological complexity,  $c,$ and $\lambda_{\mathrm{LE}}$ scale identically and quadratically with $g$ appears to be specific to the uncorrelated case $\tau=0$.


\hyperlink{Fig3}{Fig.\ 3b} shows that the dimensionality of the dynamics as measured by the participation ratio (see, e.g., \cite{clark2023dimension}) depends on  $\tau$ in a similar way to the largest Lyapunov exponent,  $\lambda_{\mathrm{LE}}$.
Dimensionality increases as $\tau$ decreases down to $\tau \approx -0.7$,  but decreases for smaller values of $\tau$. Thus, when $\tau \gtrsim -0.7$, increasing symmetry by increasing $\tau$ results in a decrease in the number of fixed points, accompanied by less chaotic (as measured by  $\lambda_{\mathrm{LE}}$) and lower-dimensional dynamics. However, this correspondence breaks down for strongly anti-symmetric coupling.

The decoupling between complexity, the maximal Lyapunov exponent, and dimensionality at $\tau \lesssim -0.7$ may be explained by two complementary phenomena. First, as $\tau \to -1$, the spectrum of the connectivity matrix becomes concentrated on the imaginary axis and the dynamics become more oscillatory, consistent with the fast decay of the maximal Lyapunov exponent toward $0$. 
Second, holding $g_{\mathrm{eff}}$ fixed while decreasing $\tau$ implies that the coupling strength scales as $g \sim (1+\tau)^{-1}$ (since $g_{\mathrm{eff}}=g(1+\tau)$), and hence  diverges as $\tau \to -1$. Such a rapid growth in $g$ is expected to increase the variance of the population activity $\sigma^2 \equiv N^{-1}\sum_j x_j^2(t)$. To understand this increase, it is useful to observe the dynamical mean-field picture, in which the recurrent input to a typical neuron is approximated by a self-consistent Gaussian drive with variance $\sigma_n^2 \equiv g^2\!\left(N^{-1}\sum_j S^2(x_j(t))\right)$.
In this mean-field description, $\sigma_n^2$  drives  the chaotic fluctuations.
For $\tau=0$, $\sigma_n^2$ is determined self-consistently by the mean-field single-neuron description. When $\tau \neq 0$, the mean-field drive is no longer governed solely by $\sigma_n^2$: correlations between reciprocal weights introduce additional $\tau$-dependent terms that affect the mean-field dynamics. Thus, the total variance $\sigma^2$ cannot be inferred from $\sigma_n^2$ alone 
(see~\cite{marti2018correlations,crisanti2018path,clark2023dimension,zou2024introduction} for further details).

Despite this distinction, we find numerically that $\sigma_n$ provides a good approximation to $\sigma$, which makes it useful for interpreting the observed variance increase at fixed $g_{\mathrm{eff}}$.
In particular, the typical Euclidean distance of trajectories from the origin, as measured by $\sqrt{N}\,\sigma$, depends only weakly on $\tau$ for $\tau \ge -0.6$. By contrast, for $\tau<-0.6$ the activity variance $\sigma^2$ increases sharply (\hyperlink{Fig4}{Fig.\ 4a}). This crossover is also reflected in the dynamical mean-field noise variance, $\sigma^2_n$, which closely tracks the rapid growth of $g$ (inset of \hyperlink{Fig4}{Fig.\ 4a}). As discussed previously, once activity moves far from the origin, the link between fixed-point structure and dynamical observables is no longer expected to be direct.

\begin{figure}\hypertarget{Fig3}{}
    \centering \vspace{0.0cm}

    \begin{subfigure}[t]{0.01\textwidth} 
        \vspace{-4.0cm}     
        \hspace*{0.45cm}    
        \textbf{(a)}
    \end{subfigure}%
    \includegraphics[width=0.23\textwidth,trim=0.0cm 0pt 0.0cm 0.0cm,clip]{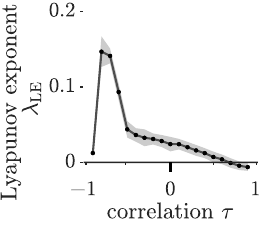}%
    \hfill 
    \begin{subfigure}[t]{0.01\textwidth} 
        \vspace{-4.0cm}     
        \hspace*{0.45cm}    
        \textbf{(b)}
    \end{subfigure}%
    \includegraphics[width=0.21\textwidth,trim=-0.0cm 0pt 0.0cm 0.0cm,clip]{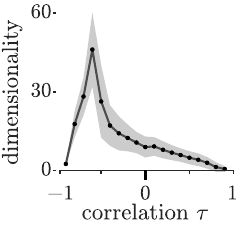}    

    \caption{(a) The maximal Lyapunov exponent as a function of $\tau$. (b) Participation dimension as a function of $\tau$. The data for these plots were obtained from $200$ realizations of the connectivity matrix ($N=2000$, $g_{\mathrm{eff}}=1.4$); gray areas indicate standard errors.}
    \label{fig:fig3}
\end{figure}

The sharp increase in the variance of the activity means that an increasing fraction of trajectories spend much of their time in regions of phase space where their nonlinear activation function, $\tanh(x),$ saturates. Since the Jacobian matrix elements are $J_{ij}=-\delta_{ij}+w_{ij}\left(1-\tanh^2(x_j)\right)$, saturation suppresses the sensitivity $S'(x_j) =(1-\tanh^2(x_j))$.
In this regime, the Jacobian approaches the negative identity matrix, so the local dynamics become predominantly contracting: small perturbations decay and local expansion rates are suppressed.
 As a result, the dimensionality contracts and the Lyapunov exponent decreases. Consistent with this picture, \hyperlink{Fig4}{Fig.\ 4b} shows that the decoupling coincides with a sharp decrease in the mean sensitivity $N^{-1}\sum_jS'(x_j)$. Together, the behavior of the participation ratio and $\lambda_{\mathrm{LE}}$ shows that while complexity shapes phase space, its impact on dynamical observables can be more complicated.

\begin{figure}\hypertarget{Fig4}{}
    \centering \vspace{0.0cm}

    \begin{subfigure}[t]{0.01\textwidth} 
        \vspace{-4.0cm}     
        \hspace*{0.45cm}    
        \textbf{(a)}
    \end{subfigure}%
    \includegraphics[width=0.21\textwidth,trim=0.0cm 0pt 0.0cm 0.0cm,clip]{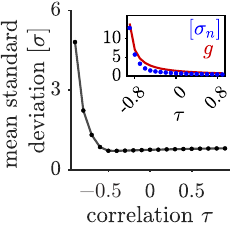}%
    \hfill 
    \begin{subfigure}[t]{0.01\textwidth} 
        \vspace{-4.0cm}     
        \hspace*{0.45cm}    
        \textbf{(b)}
    \end{subfigure}%
    \includegraphics[width=0.21\textwidth,trim=-0.0cm 0pt 0.0cm 0.0cm,clip]{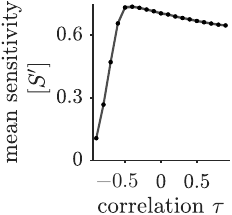}    
    
     \caption{
(a) The time-averaged standard deviation of the activity $[\sigma]$ as a function of $\tau$. Inset shows the noise standard deviation $[\sigma_{\mathrm{n}}]$ (blue) and coupling strength $g$ (red). (b) The mean sensitivity $[S']$ (averaged over network size) as a function of $\tau$. In both panels, each data point is obtained by averaging over $200$ realizations of the connectivity matrix with $N=2000$ and $g_{\mathrm{eff}}=1.4$; shaded regions denote standard errors (mostly smaller than marker size). Square brackets $[\cdots]$ denote time averages over the respective dynamics of each realization.}
    \label{fig:fig4}
\end{figure}
It is notoriously difficult to reliably estimate the number of fixed points numerically near onset. We therefore use the transient path length as a geometric proxy that is easier to measure in this regime.  Near onset, we argue that fixed points act as local attractors, saddles, or repellers that organize trajectories and shape phase-space flow. In this picture, changes in the abundance of fixed points can be reflected in the extent of transient excursions, quantified by the cumulative distance traveled along a trajectory. Such excursions can also be influenced by basin geometry and by how fixed points are distributed in phase space. 

To test the hypothesis that changes in fixed-point number near onset are reflected in the transient behavior, we compute the total path length per neuron, $\mathcal{L} :=  N^{-1/2} \int_0^{t_\text{fp}} \lVert{\dot{\mathbf{x}}(t)}\rVert   \mathrm{d}t $,  where $t_\text{fp}$ is the time of convergence to a stable fixed-point solution. Empirically, for fixed $g_{\mathrm{eff}}$ and $\tau>0$ (whenever stable fixed points were observed), the fixed-point distance from the origin depends only weakly on $\tau$ (\hyperlink{FigS1}{Fig.\ A1}), so changes in $\mathcal{L}$ primarily reflect the trajectory length until convergence, rather than merely changes in the final fixed-point magnitude. Consistent with the decrease in complexity with increasing symmetry, $\mathcal{L}$ decreases as $\tau$ increases (\hyperlink{Fig5}{Fig.\ 5a}). Moreover, if the expected number of fixed points decreases with $\tau$, we expect the system-size dependence of $\mathcal{L}$ to weaken: \hyperlink{Fig5}{Fig.\ 5b} shows that $\mathcal{L}$ grows more slowly with $N$ as $\tau$ increases, suggesting a continued monotonic decrease of complexity with $\tau$  for $\tau>0$. We do not report path-lengths of networks with $\tau<0$ due to the low probability of observing stable fixed-point solutions even for small $N$.

\begin{figure}\hypertarget{Fig5}{}
    \centering \vspace{0.0cm}

    \begin{subfigure}[t]{0.01\textwidth} 
        \vspace{-4.0cm}     
        \hspace*{0.45cm}    
        \textbf{(a)}
    \end{subfigure}%
    \includegraphics[width=0.21\textwidth,trim=0.0cm 0pt 0.0cm 0.0cm,clip]{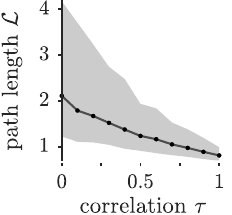}%
    \hfill 
    \begin{subfigure}[t]{0.01\textwidth} 
        \vspace{-4.0cm}     
        \hspace*{0.45cm}    
        \textbf{(b)}
    \end{subfigure}%
    \includegraphics[width=0.21\textwidth,trim=-0.0cm 0pt 0.0cm 0.0cm,clip]{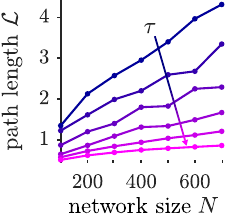}      
     \caption{(a) Median path length $\mathcal{L}$ as a function of $\tau$ for $N=700$; shading indicates the interquartile range (spanning first to third quartiles). (b) Average path length $\mathcal{L}$ versus $N$ for different values of $\tau\in\{0,0.2,0.4,0.6,0.8,1\}$, ordered top to bottom as $\tau$ increases: $\tau=0$ (solid blue, top) to $\tau=1$ (pink, bottom), shown by the diagonal arrow. In both panels, $g_{\mathrm{eff}}=1.2$.}
    \label{fig:fig5}
\end{figure}

Partially symmetric connectivity is a prominent cortical motif and has been linked to characteristic patterns of circuit dynamics and function \cite{markram1997physiology,song2005highly,ko2011functional,cossell2015functional,hart2020recurrent}. Yet it remains unclear how partial symmetry and anti-symmetry reshape the fixed point landscape and how this landscape relates to network dynamics. Here we showed that the topological complexity of partially symmetric RNNs depends non-trivially on coupling strength and reciprocity. For $g_{\mathrm{eff}}>1$ and anti-symmetric connectivity ($\tau<0$), the number of fixed points grows exponentially with $N$ with a rate that increases linearly with $g_{\mathrm{eff}}$ and whose prefactor changes sign precisely for uncorrelated connectivities, accounting for the quadratic scaling previously observed in uncorrelated ensembles. In contrast, for partially symmetric networks our leading-order estimate becomes negative near onset within our approximation. Complexity and dynamics tend to co-vary, with decreasing complexity accompanied by reduced dimensionality and a smaller maximal Lyapunov exponent, although in the strongly anti-symmetric regime this correspondence breaks down. Related non-monotonic trends in the dynamics also appear under higher-order cyclic correlations \cite{sherf2025chaos}, where weak negative correlations maximize chaotic activity while stronger negative correlations promote oscillations and reduced effective dimensionality.

These findings complement recent proposals that partial symmetry can slow fluctuations through marginal stability \cite{berlemont2022glassy}. Together, they imply that reciprocity modulates both the complexity of phase space and the stability of fixed points, while the dynamics  also depends on how those fixed points are distributed in phase space.

From a computational perspective, complexity relates to both memory and learning. In attractor-based models (e.g., Hopfield networks), memory capacity is tied to the number of stable fixed points. However, even when stable fixed points constitute only a fraction of the total number of fixed points, as could be the case in finite-size partially symmetric RNNs, the memory capacity is governed by the subset of stable or marginally stable attractors and their basins \cite{storkey1999basins,rolls2010attractor}.
 Complexity can also influence learning through trajectory geometry. For a linear readout $y(t)=w^\top x(t)$, where $w$ denotes the output weights, the output total variation obeys
\begin{equation*}
\mathrm{TV}(y) := \int_0^T |\dot y(t)|\,\mathrm{d}t
\;\le\; \|w\|\int_0^T \|\dot x(t)\|\,\mathrm{d}t \propto\|w\| \mathcal{L},
\end{equation*}
so longer path lengths allow greater output modulation for fixed $\|w\|$, and can improve linear separability in practice. Trajectory length may also affect biological plasticity by shaping pre- and postsynaptic spike-timing statistics, and thus the opportunity for spike-timing--dependent updates \cite{bi1998synaptic,caporale2008spike,sherf2025synaptic}.

Several directions follow naturally from this work. On the theory side, an informative next step is to derive the quenched complexity and to characterize the distribution of the fixed points and their stability indices. It would also be valuable to extend this analysis to higher-order correlations to test whether symmetry and positive correlations generically suppress complexity in structured disordered networks. Our theory also motivates biological tests: across circuits or cell types with different degrees of reciprocal coupling, one can ask whether oscillatory dynamics and the effective dimensionality of population activity change in a consistent way. Additionally, one can ask whether reciprocal connectivity varies systematically across cortical areas in the way that is related to the observed dynamics and computations. For example, one can ask whether regions supporting strong working memory (e.g., primate prefrontal cortex \cite{funahashi2017working}) exhibit different reciprocity than primary sensory cortex.

Our work thus establishes a quantitative, testable relationship between reciprocal connectivity and neural dynamics. We also show that topological complexity relates to observable dynamical properties, and thus offers a tool for probing more general connections between the  structure of neural circuits, their dynamics and function.
More broadly, our theory supports the view that complexity depends strongly on the statistical structure of interactions, not only on overall coupling strength. This observation is consistent with those made for other disordered systems, including random ecological networks \cite{ros2023quenched}, random evolutionary games \cite{duong2019expected}, and spin glasses \cite{mckenna2024complexity}. Together, these examples suggest a broader link between the complexity of high-dimensional disordered systems and the structure of interactions among their constituent elements.

\begin{acknowledgments}
This research was supported in part by grants from NSF NeuroNex (DBI-1707400), NIH (R01 MH130416) to NS, XP and KJ, by NSF LEAPS grant (DMS-2137614) to ST, by NSF (DMS-2207647) and NSF (DMS-2235451)  to KJ, by the NSF and DoD OUSD (R \& E) under Cooperative Agreement PHY-2229929 (The NSF AI Institute for Artificial and Natural Intelligence, ARNI) to XP,
and by the Simons Foundation (MPS-NITMB-00005320) through the NSF-Simons National Institute for Theory and Mathematics in Biology to NS, XP, KJ, and KEB. This work was completed in part with resources provided by the Research Computing Data Core at the University of Houston.
\end{acknowledgments}

\begin{appendix}

\renewcommand{\thefigure}{A\arabic{figure}}
\setcounter{figure}{0}

\renewcommand{\thetable}{A\arabic{table}}
\setcounter{table}{0}

\renewcommand{\theequation}{A\arabic{equation}}
\setcounter{equation}{0}

\section{Spectrum of $\mathbf W\cdot \Delta( S'(\mathbf x^\ast))$\label{sec:spectrumWD}}
In this appendix, we show that the largest real part of the eigenvalues of $\mathbf W \cdot \Delta(S'(\mathbf x^\ast))$ is bounded by $g(1+\tau)$. 
Denote the diagonal entries of $\Delta(S'(\mathbf x^\ast))$ by $d_1, d_2, \ldots, d_N$.  We have $0\le d_i\le 1$ for all $i=1,2, \ldots, N$, since $S$ is a sigmoid function with maximal derivative $S'(0)=1$. Write $\mathbf J =\mathbf W \cdot \Delta(S'(\mathbf x^\ast))$ so that $J_{ij}=W_{ij}d_j$. We have $\langle J_{ij}\rangle =d_j\langle W_{ij}\rangle =0$ and
\begin{align*}
    \langle J_{ij}J_{kl}\rangle =d_jd_l\langle W_{ij}W_{kl}\rangle =\frac{g^2}{N}(d_j^2 \delta_{ik}\delta_{jl} + \tau d_id_j\delta_{il}\delta_{jk}).
\end{align*}
The covariance structure of the matrix $\mathbf J$ has the same form as what is discussed in \cite{kuczala2016eigenvalue}, with $g_{ij}=gd_j$. 
Denote the largest real part of the eigenvalues of $\mathbf J$ by $R\ge 0$. It follows from \cite{kuczala2016eigenvalue} that $R$, together with the other $N$ variables $(c_1, \ldots, c_N)$, solve the following system of equations 
\begin{align}
    \label{eq:A1}
    &c_i = \left(R - \frac{\tau g d_i}{N}\sum_{j=1}^N c_j gd_j\right)^{-1}, \ \ i=1,2,\ldots, N\\
    \label{eq:A2}
    &\sum_{j=1}^N c_j^2 (gd_j)^2 = N.
\end{align}
Note that when $\tau=0$, then $R = g\sqrt{\frac{1}{N}\sum_{i=1}^N d_i^2}$, which recovers the ``inhomogeneous'' circular law established in \cite{wainrib2013topological}; when $d_1=d_2=\cdots=d_N\equiv d$, we have $R=gd(1+\tau)\le g(1+\tau)$, and the limiting spectrum of $\mathbf W\cdot \Delta( S'(\mathbf x^\ast))$ is a rescaled elliptical law.
It suffices to show that the solution of the above system of equations satisfies $R\le g(1+\tau)$ for any $\tau \in [-1, 1]$.  

To start, we write $Y:=\sum_{j=1}^N c_j d_j$, so that the first $N$ equations in \eqref{eq:A1} become
\[
c_i R - \frac{c_i \tau g^2 d_i}{N}Y=1, \ \ i=1,2,\ldots, N.
\]
Multiplying both sides by $c_i  d_i$ and summing over all $i=1,2,\ldots, N$, we get
\[
R \sum_{i=1}^N c_i^2 d_i - \tau  Y = Y,
\]
where we have used Eq.~\eqref{eq:A2}.
Thus, we have 
\begin{align}
\notag
R &= (1+\tau)Y \left(\sum_{i=1}^N c_i^2 d_i \right)^{-1}\\
\label{eq:A3}
&=(1+\tau) \frac{c_1d_1+\cdots+ c_N d_N}{c_1^2 d_1 + \cdots + c_N^2 d_N}.
\end{align}
Since $R\ge 0$, $\tau \ge -1$ and $d_i\ge 0$, the numerator in \eqref{eq:A3}, $Y=c_1d_1+\cdots + c_N d_N\ge 0$. 
Using Cauchy-Schwarz inequality and Eq.~\eqref{eq:A2}, we have
\begin{align*}
Y^2 = \left(\sum_{j=1}^N c_j d_j\right)^2 \le N \sum_{j=1}^Nc_j^2 d_j^2 =N^2/g^2,
\end{align*}
which implies $Y\le N/g$. Since $0\le d_i\le 1$, the denominator on the right hand side of Eq.~\eqref{eq:A3} is at least
\[
c_1^2d_1^2 + \cdots + c_N^2 d_N^2 = N/g^2.
\]
Putting all this together, we arrive at
\begin{align*}
    R \le (1+\tau) \frac{N/g}{N/g^2}=g(1+\tau),
\end{align*}
which completes the proof.

\section{Uniqueness of the trivial fixed point\label{sec:uniq-fix}}

Let $F(\mathbf x):=-\mathbf x + \mathbf W S(\mathbf x)$ denote the  right side of Eq.~\eqref{eq:nn}. We show in this section that when $g_{\text{eff}}<1$, if for any fixed point $\mathbf x^\ast$ of the system in Eq.~\eqref{eq:nn}, all  eigenvalues of the Jacobian matrix  
$$DF(\mathbf x^\ast):=-\mathbf I +\mathbf W\cdot \Delta(S'(\mathbf x^\ast))=-\mathbf I + \mathbf J$$ 
have negative real part, then the trivial fixed point, $\mathbf x^\ast = \mathbf 0$ is the only fixed point. We will make use of the Brouwer degree of the map $F(\mathbf x):=-\mathbf x + \mathbf W S(\mathbf x)$ at $\mathbf 0$, defined as
\begin{align*}
\deg(F,\mathbf 0) &:= \sum_{\mathbf y\in F^{-1}(\mathbf 0)}\text{sign}(\det(DF(\mathbf y)))\\
&= \sum_{\mathbf x^\ast}\text{sign}(\det(DF(\mathbf x^\ast))).
\end{align*}
First of all, since all eigenvalues of $DF(\mathbf x^\ast)$ have negative real parts, we have
$\text{sign}(\det(DF(\mathbf x^\ast)))=(-1)^N$
and 
\begin{align*}
\deg(F,\mathbf 0) &=A_N(g, \tau)\cdot (-1)^N.
\end{align*}
On the other hand, consider the map $G(\mathbf x):=-\mathbf x + g_{\text{eff}} \mathbf x$, which has a Jacobian $DG(\mathbf x)=(g_{\text{eff}}-1)\mathbf I$. One can easily compute that when $g_{\text{eff}}<1$, the Brouwer degree of $G$ is
\[
\deg(G,\mathbf 0) =(-1)^N.
\]
It is easy to check that the interpolated map $H(t, \mathbf x) := tF(\mathbf x) + 
(1-t)G(\mathbf x)$  has $\mathbf 0$ as a regular point for all $t\in[0,1]$, which implies that 
\[
\deg(F,\mathbf 0) = \deg(G,\mathbf 0)=(-1)^N
\]
by the homotopical invariance of the Brouwer degree. It follows that $A_N(g,\tau)=1$, i.e., there is one single fixed point $\mathbf x^\ast=\mathbf 0$ for the system given in \cref{eq:nn}.

\section{Computation of the topological complexity\label{sec:topo-complex}}
\setcounter{equation}{3}
Recall that when $g_{\text{eff}}>1$ and as $g_{\text{eff}}\to 1^+$, the topological complexity is asymptotically $c(g,\tau)=\int_\mathbb C \log|-1+z|\,   \rho_{g, \tau}(z)\mathrm{d}z$, where $\rho_{g, \tau}(z)$ is the density of the  uniform distribution on the ellipse $\mathcal E_{ab}$:
\[
\rho_{g, \tau}(z)=\frac{\mathbbm 1\{z\in \mathcal E_{ab}\}}{\pi g^2(1-\tau^2)}
\]

In this section, we show that
\begin{align}
\label{eqn:integral-over-ellipse}
      c(g,\tau) = \frac{1}{2 g^2 (1 + \tau)} + \log g - \frac{1}{2},
\end{align}
Firstly, since $\Re (\log(\omega)) = \log|\omega|$ for $\omega \in \mathbb{C}$ in the domain of $\log| \cdot |$,  we write $z=x+iy$ and have
\begin{align*}
\int_{\mathbb{C}}^{}& \log |z-t| \rho_{g, \tau}(z)\mathrm{d} z\\
&=  
\frac{1}{\pi g^2 (1 - \tau^2)} \Re \int_{{x+iy \in \mathcal E}_{ab}}^{} \log (x + iy - t)\mathrm{d}x \mathrm{d} y
\end{align*}
for any $t>0$. We next make the following change of variables: 
$$\left\{ \begin{array}{cl}
x = g (1 + \tau)r \cos (\theta) \\
y = g (1 - \tau)r \sin (\theta), \\
\end{array} \right.$$
where $ 0 \leq r \leq 1$ and $ 0 \leq \theta \leq 2 \pi.$ The Jacobian from the change of variables is then 

\begin{align*}
\left | \frac{\partial (x,y)}{\partial(r, \theta)} \right | &= g^2 (1- \tau^2) \begin{vmatrix}
\cos (\theta) & -r \sin (\theta)  \\
\sin(\theta) & r \cos(\theta)
\end{vmatrix} \\
 &= g^2(1 - \tau^2) r.
\end{align*}

\noindent Then, we have that 
\begin{equation*}
\begin{split}
&\frac{\mathrm{d}}{\mathrm{d}t} \int_{\mathbb{C}}^{} \Re \log (z-t) \mathrm{d}\rho_{g, \tau}(z)  = 
\int_{\mathbb{C}}^{}\frac{\mathrm{d}}{\mathrm{d}t} \Re \log (z-t) \mathrm{d}\rho_{g, \tau}(z)= \\
&  \frac{1}{\pi} \Re \int_{0}^{1}r\mathrm{d}r \int_{0}^{2\pi} \frac{\mathrm{d}\theta}{g(1+\tau)r\cos(\theta) + i g(1-\tau)r \sin (\theta) -t}.
\end{split}
\end{equation*}

\noindent To evaluate the inner integral, we make use of the contour integration. Let $z = e^{i \theta}.$ Then, $\mathrm{d} \theta = \frac{\mathrm{d}z}{iz}.$ With 
$$\cos(\theta) = \frac{z + z^{-1}}{2} \ \ \text{and} \ \ \sin(\theta) = \frac{z - z^{-1}}{2i},$$
we have 
$$g(1+\tau)r\cos(\theta) + i g(1-\tau)r \sin (\theta) - t = g r (z + \tau z^{-1}) - t.$$

\noindent The inner integral now becomes a contour integral over the unit circle $|z| = 1,$ and reduces to  
$$\frac{1}{i}\oint_{|z| =1}^{}\frac{\mathrm{d}z}{g r (z + \tau z^{-1}) - t}.$$
Let $f(z) = g r (z + \tau z^{-1}) - t.$ To find the poles, we solve $f(z) =0$ and obtain
$$z_{\pm} = \frac{t \pm \sqrt{t^2 - 4 g^2 r^2 \tau}}{2 g r}.$$ 

\noindent This implies that, by the Residue Theorem,

\begin{equation*}
\begin{split}
\frac{1}{i}\oint_{|z| =1}^{}\frac{\mathrm{d}z}{g r (z + \tau z^{-1}) - t} &= \frac{2 \pi i \text{Res} (z_{-})}{i} \\
&= \frac{2 \pi}{\sqrt{t^2 - 4 g^2 r^2 \tau}} \ \ 
\end{split}
\end{equation*}
$\text{when} \ \  g r (1 + \tau) < t,$
and $0$ otherwise. If $0 < t < g ( 1+ \tau),$ (which is the case we are interested in), then this implies the inner integral only contributes when $r < \frac{t}{g (1 + \tau).}$ Thus, under these conditions, 
$$ \frac{\mathrm{d}}{\mathrm{d}t}\int_{\mathbb{C}}^{} \log |z-t| \mathrm{d}\rho_{g, \tau}(z) = \frac{1}{\pi} \Re \int_{0}^{\frac{t}{g (1 + \tau)}}\frac{2 \pi}{\sqrt{t^2 - 4 g^2 r^2 \tau}} r\mathrm{d}r .$$
Let $r^* = \frac{t}{g (1 + \tau)}$ and let $c^2 = 4g^2 \tau.$ Substituting $u = r^2 \implies \mathrm{d}u = 2 r \mathrm{d} r,$ then 
$$\frac{1}{\pi} \Re \int_{0}^{\frac{t}{g (1 + \tau)}}\frac{2 \pi}{\sqrt{t^2 - 4 g^2 r^2 \tau}} r\mathrm{d}r = \frac{1}{2}\int_{0}^{r^{*2}} \frac{\mathrm{d}u}{\sqrt{t^2 - c^2u}}.$$
We now must make another substitution $v = t^2 - c^2u \implies \mathrm{d}v = -c^2 \mathrm{d}u,$ which leads us to 
$$\frac{1}{c^2}\int_{t^2 - c^2 r^{*2}}^{t^2} v^{-1/2} \mathrm{d} v.$$
After a lot of algebra, we finally get the result 
$$\frac{\mathrm{d}}{\mathrm{d}t}\int_{\mathbb{C}}^{} \log |z-t| \mathrm{d}\rho_{g, \tau}(z) = \frac{t}{g^2 ( 1 + \tau)},$$ and integrating and setting $t = 1$ we have that 
$$\int_{\mathbb{C}}^{} \log |z-1| \mathrm{d}\rho_{g, \tau}(z) = \frac{1}{2 g^2 (1 + \tau)} + C.$$

Finally, using the fact that when $\tau = 0$ the integral is equal to $\frac{1}{2 g^2} + \log g - \frac{1}{2}$,  as computed in \cite{wainrib2013topological}, we determine the integration constant $C = \log g - \frac{1}{2}$, which finishes the computation of the topological complexity.


\vskip5pt

\section{Distance of stable fixed points}
When stable fixed points are found, their distance depends only weakly on the symmetry level of the connectivity matrix $\tau$. The distance of a single fixed point is measured as $\sqrt{N} \sigma$ where $\sigma$ is the standard deviation of the coordinates of the fixed point. In \hyperlink{FigS1}{Fig.\ A1}, we show $\sigma$ as a function of $\tau$, for $N=700$. The variations of $\sigma$ are on the order of $10^{-2}$, where $\sigma$ shows very slow decrease with an increasing value of $\tau$.
\begin{figure}[ht]\hypertarget{FigS1}{}
    \centering
    \includegraphics[width=0.22\textwidth]{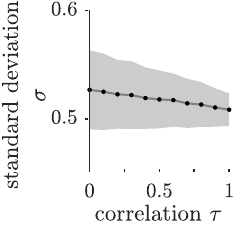}
    \caption{The normalized distance $\sigma$ as a function of $\tau$, for $N=700$ for $g_{\mathrm{eff}}=1.2$. The gray area shows the standard errors.}
\end{figure} 
The convergence to a fixed point solution occurs at an increasing frequency with $\tau$. Thus, the dynamics in $740/3000$ realizations with $\tau=0$ have converged to a fixed point solution. For $\tau=1$, the dynamics in $2960/3000$ realizations have converged to a fixed point solution within the allotted time. 

\end{appendix}

\bibliography{MainPaper}

\end{document}